\documentclass[twocolumn]{article}
\usepackage{graphicx}
\usepackage{authblk}
\usepackage[font={small}]{caption} %James: Not immediatley clear to me that these caption-related packages are necessary
\usepackage{csquotes}
\usepackage{amsmath}
\usepackage{microtype}
\usepackage{tabu}
\usepackage{longtable}
\usepackage{subfig}
\usepackage{setspace}
\usepackage{array}
\usepackage{booktabs}
\usepackage{verbatim}
\usepackage{comment}
\usepackage{multirow}
\usepackage{algorithm} %gives floating algorithm environment, like figure* and table
\usepackage{algorithmic} %old but reliable; there are others
\usepackage{epsfig}
\usepackage{url}

\graphicspath{{plots/}} 
\title{Automatic Rain and Cicada Chorus Filtering of Bird Acoustic Data}

\author[1]{Alexander Brown}
\author[1]{Saurabh Garg}
\author[1]{James Montgomery}
\affil[1]{School of Technology, Environments and Design, University of Tasmania, Hobart, Tasmania, Australia}

\begin{document}

\maketitle

%\linenumbers
\begin{abstract}
	%Automatic recording and analysis of bird calls is becoming an important way to understand changes in bird populations and assess environmental health. An issue currently proving problematic with the automatic analysis of bird recordings is interference from noise that can mask vocalisations of interest. As such, noise reduction can greatly increase the accuracy of automatic analyses. 
	
	Recording and analysing environmental audio recordings has become a common approach for monitoring the environment. A current problem with performing analyses of environmental recordings is interference from noise that can mask sounds of interest. This makes detecting these sounds more difficult and can require additional resources. While some work has been done to remove stationary noise from environmental recordings, there has been little effort to remove noise from non-stationary sources, such as rain, wind, engines, and animal vocalisations that are not of interest.
	
	In this paper, we address the challenge of filtering noise from rain and cicada choruses from recordings containing bird sound. We improve upon previously established classification approaches using acoustic indices and Mel Frequency Cepstral Coefficients (MFCCs) as acoustic features to detect these noise sources, approaching the problem with the motivation of removing these sounds. We investigate the use of acoustic indices, and machine learning classifiers to find the most effective filters. The approach we use enables users to set thresholds to increase or decrease the sensitivity of classification, based on the prediction probability outputted by classifiers. We also propose a novel approach to remove cicada choruses using band-pass filters
	
	Our threshold-based approach (Random Forest with Acoustic Indices and Mel Frequency Cepstral Coefficients (MFCCs)) for rain detection achieves an AUC of 0.9881 and is more accurate than existing approaches when set to the same sensitivities. We also detect cicada choruses in our training set with 100\% accuracy using 10-folds cross validation. Our cicada filtering approach greatly increased the median signal to noise ratios of affected recordings from 0.53 for unfiltered audio to 1.86 to audio filtered by both the cicada filter and a stationary noise filter.

\end{abstract}

\section{Introduction}
\label{ch:targeted}

%Bird monitoring has recently been of great research interest because of its broad range of applications including tracking bird migration \cite{Stepanian_2016}, monitoring biodiversity \cite{Salamon_2016}, and detecting endangered species \cite{Willacy_2015}. Traditionally, to monitor animals, experts need to be present in the region of interest \cite{Alonso_2017}. However, this is time consuming and expensive, and is impractical for large scale studies. A more efficient approach involves using sensors to record distinct animal vocalisations. However, with the large amount of recording data necessary to monitor an ecosystem, it is impractical for humans to listen and manually identify birds from these recordings. Hence, researchers have turned to automated techniques to process these environmental recordings. The science of analysing animal vocalisations is called bioacoustics.

Recently, there has been research interest in designing processes to detect and monitor animal species using unattended environmental recordings \cite{Ganchev_2017}. A key challenge in achieving this goal is interference from noise which can mask vocalisations of interest \cite{Alonso_2017,Stowell_2016}. This makes vocalisations more difficult or impossible to detect, and can result in wasting resources on examining audio that cannot be feasibly analysed.

While some work has gone into filtering stationary (i.e. constant) background noise for both speech \cite{Boll_1979,Ephraim_1984} and bioacoustics processing \cite{Ren_2008,Priyadarshani_2016}, there are many sounds, such as rain, wind, car sounds, and animals that are not of interest, that cannot be removed using these filters. In this work, we examine and evaluate techniques to filter noise from rain and cicada choruses for use in a theoretical bird sound analysis. These sources are common in the environmental recordings being analysed, interfere significantly with signals of interest, and have distinct characteristics that can help in detecting them \cite{Ferroudj_2014,Towsey_2014a}. 

There has been limited research into removing these noise sources, and the research that has been done has focussed on detecting these sources, though not for the purposes of removal \cite{Ferroudj_2014,Towsey_2014a,Bedoya_2017}. Only the presence or raw intensity of noise from some sources of environmental noise, such as rain, have been considered rather than the extent of which noise interferes with any signal in the recording. For example, light rain might interfere significantly with a quiet animal call, but a loud animal might be very clear even in the presence of heavy rain. Furthermore, they do not consider that users might desire different to filter noise sources with different sensitivities. For example, a researcher might only want to keep the cleanest samples possible, but another researcher might want to keep everything that might contain a vocalisation of interest.

To address current limitations, we propose two filters: one for cicada choruses and another for rain. These filters utilise multiple  acoustic indices in combination with MFCCs and investigated a much wider set of machine learning classification approaches in contrast to previous research \cite{Ferroudj_2014,Towsey_2014a}. We determine the most effective filtering configuration by evaluating the ability of combinations of acoustic features, classifiers, and other filters to detect rain and cicada choruses, using Area Under the Receiver Operating Characteristic Curve (AUC) as our primary metric. We compare feature sets, classifiers, and the effect of other filters deeper than previous work and this results in more accurate classification and filtering. While samples classified as containing rain are removed, we introduce a second step to filtering cicada choruses which removes only the frequency range containing the choruses.

The filters are designed to work with thresholds based on the probabilities of samples containing the noise source of interest (i.e. rain or cicada chorus). This allows users to determine the sensitivities of the filters. They are also trained to classify samples based on how much they interfere with sounds of interest, which are more suited for filtering than intensity-based classifiers in previous work.

In the next section, we discuss current works into filtering rain and cicada sounds and their limitations. We then present a methodology for cicada chorus and rain detection in Section \ref{sec:methodology}. Then, in Section \ref{sec:raincicadaresults}, we present the results of the rain and cicada detection. In Section \ref{sec:cicRemoval}, we propose a filter to remove cicada choruses from environmental recordings and evaluate its effectiveness. Finally, we conclude our work and propose future directions in Section \ref{sec:conclusion}.

\begin{comment}
Rain and cicada detection use similar approaches and so are discussed here together. This involves training a C4.5 classifier \cite{Quinlan_1993} using several acoustic indices. A methodology is then specified in Section \ref{sec:raincicadameth} before the results are given in Section \ref{sec:raincicadaresults}.

The techniques here are very similar to those used in existing literature, particularly \cite{Ferroudj_2015}. This work does not improve significantly on existing work, but combines these approaches with other noise reduction approaches to give a pipeline capable of efficiently removing many types of noise at once.

\end{comment}

\section{Related Work}
\label{sec:nf}
Current work into processing rain and cicada sounds has been limited to detection approaches. We examine these existing approaches, and how they could be improved upon to detect more accurately and to remove these sounds.

\subsection{Rain Detection}

\begin{figure}[h]
	\centering
	\includegraphics[width=1\linewidth]{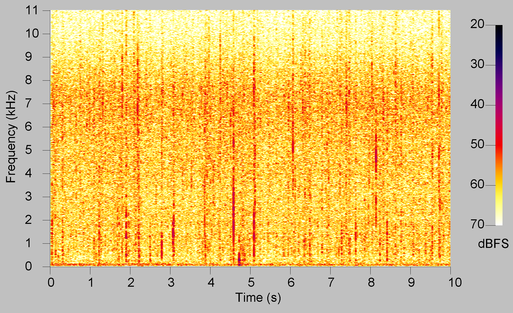}
	\caption[Spectrogram of 10 seconds of audio containing rain]{Spectrogram of 10 seconds of audio containing rain. Note the vertical bands that cover the whole spectrogram representing rain drops.}
	\label{fig:rainspectrogram}
\end{figure}

Although classification approaches are widely used in bioacoustic analyses, approaches used to classify animal species such as birds might not be well suited for classifying environmental sounds such as rain. For example, Mel Frequency Cepstral Coefficients (MFCCs) have been successfully used as a feature set for many bioacoustic classifiation tasks \cite{Alonso_2017,Ness_2013,Stowell_2014,Mirzaei_2012}. However, rain sound has different characteristics to animal sounds which MFCCs are not well suited to identifying. They are not effective in classifying noise-like signals \cite{Chu_2009} and signals with narrow spectral peaks \cite{Chu_2008}, which are characteristic of rain (see Figure \ref{fig:rainspectrogram}).

As such, alternative feature sets have been used to detect rain. Ferroudj et al. \cite{Ferroudj_2014} propose to use a set of five ``acoustic indices"

\begin{itemize}
	\item \textbf{Acoustic Complexity Index (ACI)} \cite{Pieretti_2011}: a measure of intensity variability 
	\item \textbf{Spectral and Temporal Entropy} \cite{Sueur_2008}: a measure of the dispersion of energy with frequency and time respectively
	\item \textbf{Background Noise (BGN)} \cite{Towsey_2013, Lamel_1981}: a measure of the amount of Gaussian distributed background noise
	\item \textbf{Spectral Cover} \cite{Towsey_2014a}: a measure of the fraction of cells in a spectrogram where spectral amplitude exceeds an empirically derived threshold.
\end{itemize}

These indices describe more general, statistical features of the audio and some of these can more accurately measure changes in intensity compared to MFCCs \cite{Towsey_2014a}. As such, these are plausibly a better feature set than MFCCs for rain classification. More information on these indices is given in Section \ref{Indices}.

In Ferroudj et al's. work, a dataset was trained using these acoustic indices as a feature set using the tree-based C4.5 classification algorithm \cite{Quinlan_1993} citing its fast learning speed, and explicit rule set as reasons for this choice. They found that this classifier was more accurate than other classifiers they tested. Using this classifier with the selected feature set, they achieved 93\% accuracy in classifying heavy rain from non-heavy rain.

Bedoya et al. \cite{Bedoya_2017} estimated the power spectral density \cite{Welch_1967} and signal to noise ratio in the frequency band where rain was found to be the most prominent, 600\textendash1200 Hz. These were compared to two thresholds. If both the spectral power density and signal to noise ratio exceeded their respective thresholds, audio was classified as rain. They also noted that rainfall sound is prominent in the 4400\textendash5600 Hz frequency range, though the prominence of light rain in this range is relatively weak and there is a potential for animal calls to be misclassified as rain. This observation does not appear to hold true in our set. A spectrogram from a sample containing rain in our data, shown in Figure \ref{fig:rainspectrogram}, suggests that rain is prominent across the entire frequency spectrum, though most prominent between 6--8 kHz. Bedoya et al. achieved a 95.23\% correlation between a pluviometer (a rain gauge) and an overall accuracy of 92.90\%. However, this accuracy increased to 99.98\% when only moderate or greater rain intensities were considered. 

Bedoya et al. \cite{Bedoya_2017} and Ferroudj et al. \cite{Ferroudj_2014} use approaches that are in some sense, mutually compatible. The SNR and PSD used by Bedoya et al. can be used as acoustic indices, adding to the set already used by Ferroudj et al. As such, it is possible to combine the two approaches and test them together to examine their combined effectiveness. However, there is no guarantee that adding extra features will generate more accurate results in general \cite{Chu_2009}, so this needs to be evaluated against the separate feature sets.

\begin{comment}
Hetherington and Groves \cite{Hetherington_2011} detect rain by looking at frames of audio with significantly more energy than frames adjacent to them for a wide range of frequency bands. If these bands fit a model resembling a straight line, the audio is considered rain. The approach was not tested in its documentation, and it is only described vaguely, which makes it difficult to implement and determine its accuracy.
\end{comment}

\subsection{Cicada Detection}

In the context of this research, any animal sound that is not from a bird is considered noise. Cicadas are particularly loud and are prominent in some environmental recordings \cite{Towsey_2014a}. This makes them a good target for noise removal.

To detect cicadas, Towsey et al. \cite{Towsey_2014a} trained a See5 (i.e. C5.0) classifier \cite{Quinlan_1993} using six acoustic indices (background noise, SNR, acoustic activity, mid-band activity, temporal entropy, and spectral entropy; see Section \ref{Indices}), to derive a simple rule by selecting a very simple tree as generated by the classifier (10 trees were derived from the data overall). This rule used two acoustic indices: spectral entropy and background noise.

The rule derived indicated that if the spectral entropy of a sample is below a threshold, and the background noise is above a threshold, then the sound was generated by cicadas. This is reflective of cicadas' loud sounds that are concentrated in narrow frequency ranges. However, the average accuracy of all trees derived was a relatively low 76\%. It is also noted by Towsey et al. \cite{Towsey_2014a} that cicada choruses decrease the SNR of the audio.

Ferroudj et al.\cite{Ferroudj_2014} applied the same technique used to detect heavy rain to detect cicadas, using acoustic indices as features for a classifier. This is also the same technique as Towsey et al., but with the feature set from their rain classifier. In a four-class problem (bird sounds, heavy rain, cicadas, and other), an accuracy of 81\% was achieved, with an 81\% recall.

\subsection{Limitations and Our Contribution}

While research has already been conducted to examine sources of noise that are targeted for removal by the present work, it has has focussed on the detection of these sounds, not their removal. The emphasis here on removing these sounds changes which sounds we want to detect, particularly from a rain detection perspective. Recordings with light rain drops could contain clean calls which could be useful for further analysis, while recordings containing heavier rain are likely to damage signals too significantly to be useful and can be removed. It is unclear whether users want to keep recordings with light rain or not. Additionally, researchers are more likely to want to keep recordings with light rain but prominent bird calls over recordings with light rain but faint calls, as prominent calls will be less damaged.

Additionally, researchers offer no approach to filter these environmental noises outside of removing entire samples. While we do not propose an improvement to this from a rain filtering perspective, it is possible to remove cicada choruses. It is possible to detect and filter these, because they occupy a narrow frequency band which can be removed without damaging bird signals outside the frequency range of the cicada chorus. As such, we design a novel algorithm for the removal of cicada choruses.

\section{Methodology For Targeted Sound Detection}
\label{sec:methodology}

Our approach for filtering rain and cicada choruses utilises machine learning, much like previous research. The selection of classifier, feature set, and additional pre-processing tasks play a significant role in the accuracy of our classifier. As such, we test many combinations of feature sets, classifiers, and filters using a training dataset to determine the best rain and cicada chorus filters.

\subsection{Targeted Noise Sources}
\label{sec:sounds}
Before examining approaches to detect and remove targeted sound sources, it is necessary to discuss and clarify the nature of the sound sources being examined.

\subsubsection{Cicada Sounds}
\begin{figure*}[h]
	\centering
	\subfloat[Cicada Chorus]{ \includegraphics[width=0.45\textwidth]{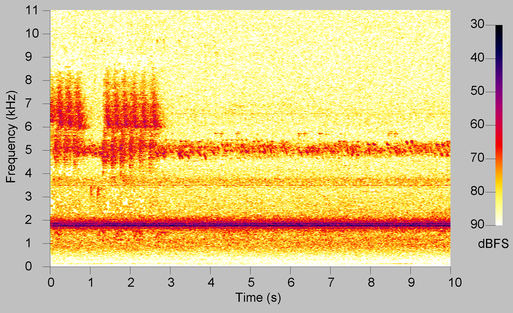} \label{cicChorus} }
	\subfloat[Cicada Chirp]{ \includegraphics[width=0.45\textwidth]{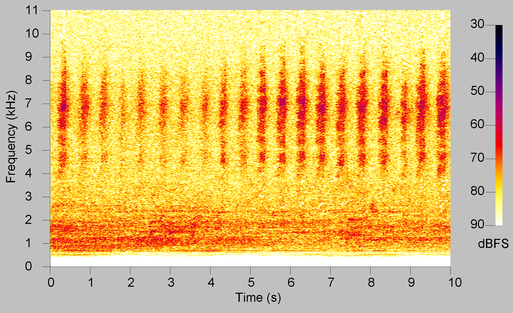} \label{cicChirp} }
	
	\caption{Spectrograms (generated by SoX \cite{SoX}) of samples with constant cicada chorus (left) (note the constant tone around 2 kHz) and noisy cicada chirps (right). Some cicada chirps have different characteristics to this spectrogram. The cicada chorus has been high-pass filtered to give a better comparison for Figure \ref{fig:cicFilter}}
	\label{cicadas}
\end{figure*}

The cicada sounds that appear in the samples can be divided into two types, which are depicted in Figure \ref{cicadas}. One is a \textit{chorus} which, relative to other sounds in the recording, has a very loud, consistent tone with a narrow frequency range. The second cicada sound type is much noisier, occupying a broad range of high frequencies (from around 6 kHz to the Nyquist frequency at 11.025 kHz for a 22.05 kHz sampled recording). These are also much less constant, with individual calls being clearly distinctive. These are made up of shorter and longer sounds. As these are short signals, they are much less likely to interfere with sounds of interest, and are more difficult to filter. As such, the focus for cicada removal here is on cicada choruses.

\subsubsection{Rain Sounds}
Rain in the sample audio set can also be divided into two types which both have significantly different characteristics. The first of these has a consistent amplitude, sounding almost like white noise. The second type has clearly audible individual rain drops, which tend to be much louder than the background noise. These are grouped together in the same class in the classification algorithm, because from a practical perspective, they will be treated the same way, but their different properties mean that the classifier will apply different rule sets for each rain type.

We define the aim of the rain classifier to be to detect any rain that is significantly louder any signal in the audio. Audio samples containing loud rain are worthless for further analysis because any desired signal that might be in the audio is too unclear to analyse further. As such, light rain is considered to be not rain for the purposes of the training data, because many signals are still clearly audible in low intensity rain. Samples containing both heavy rain and very prominent non-rain signals that can be heard over the rain are also considered to be not rain, as they contain information that is likely to be usable in later analysis. 

However, given that users might want to remove rain to different extents, we examine the classification probabilities given by the classifiers, which can then be used to develop thresholds which can vary the sensitivity of classifiers. These thresholds will not correspond to intensity, but more likely the level by which a recording is going to be absent of bird sounds that can be easily heard and processed.

\begin{figure*}[h]
	\centering
	\subfloat[Rain Drops]{ \includegraphics[width=0.45\textwidth]{"RainDropsSpectrogram"} \label{rainDrop} }
	\subfloat[Noisy Rain]{ \includegraphics[width=0.45\textwidth]{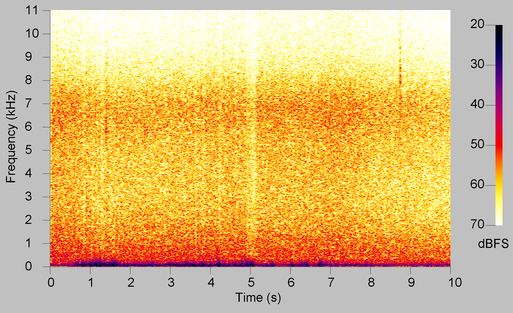} \label{rainNoise} }
	
	\caption{Comparison of the two rain types. The left spectrogram and clear vertical lines indicating clearly audible rain drops, whereas the right image has a constant noisy profile.}
	\label{rain}
\end{figure*}

\subsection{Classifiers}

We utilise machine learning classifiers to filter out rain and cicada choruses. For rain filtering, the classifier is the filter, but for cicada chorus filtering, the classifier is used to determine if a sample should undergo the band-pass filtering approach described in Section \ref{sec:cicFilterResults}. Instead of using a discrete classification, we use the probability of a classifier detecting the presence or absence of rain and cicada choruses as a basis for classification. This allows us to use the prediction probability as a threshold which can be varied to increase or decrease the sensitivity of the classifier. We compare multiple classifiers in our evaluation to which is most effective in detecting and filtering rain and cicada choruses. Classifiers tested are:

\begin{itemize}
	\item \textbf{Naive Bayes} \cite{John_1995}: Utilises Bayes' Theorem to calculate the probability that an instance belongs to a class, which in turn drives classification.
	\item \textbf{IBk} \cite{Aha_1991}: k-nearest neighbours by identifying the instances with the closest Euclidian distance.
	\item \textbf{JRIP} \cite{Cohen_1995}: Rule-based classifier that incrementally derives rules, selecting for the highest information gain.
	\item \textbf{J48 (i.e. C4.5)} \cite{Quinlan_1993} Top-down tree-based classifier.
	\item \textbf{SMO} \cite{Platt_1998,Keerthi_2001}: Support vector machine-based classifier.
	\item \textbf{Random Forest} \cite{Breiman_2001}: Generates a large number of decision trees and uses the mode classification from these tree to classify instances.
\end{itemize}

We utilised the Weka \cite{Hall_2009} implementation of each classifier using default settings, except for IBk and SMO. For IBk, values for $k$ between 1 and 25 in increments of 2 are tested, and only the best result for each configuration is given where applicable. For SMO, an option to build calibration models is enabled to properly allow prediction probabilities (and hence, thresholds).

\subsection{Acoustic Features}
\label{sec:features}

An important aspect to any classifier is to train it using the most effective feature set so that it can distinguish between classes. In this investigation, we calculate several statistical features of the audio recordings to train classifiers to determine the presence of rain and cicada choruses. These are calculated automatically from examining the amplitude and frequency components of the audio. We focus on acoustic indices and Mel Frequency Cepstral Coefficients (MFCCs) in this study. The set of features used in this study is more thorough than in previous works.

\subsubsection{Acoustic Indices}
\label{Indices}

Acoustic describe specific properties of the audio signal that can be characteristic of specific sounds, such as rain and cicada choruses. Indices used are:

\textbf{Temporal Entropy} \cite{Sueur_2008}: Measures the dispersal of signal intensity over time. Higher values indicate a more even distribution of intensity, whereas lower levels indicate a more variable level of intensity. This is calculated by using a Hilbert Transform to get the analytic signal. Here, a Short Time Fourier Transform (STFT) of the data is taken (this is used for calculating multiple indices), removing all negative frequency components, and then computing an inverse STFT. Taking the magnitude of values in this inverted transform gives the signal amplitude. This is then used to calculate the probability mass function for the amplitude over time:

\begin{equation}
A(t)=\frac{|\zeta (t)|}{\sum_{t=1}^{n}|\zeta (t)|} \text{such that} \sum_{t=1}^{n}A(t)=1
\end{equation}
Where $t$ is time (in terms of the sample number), $n$ is the total number of samples, and $\zeta(t)$ is the analytic signal of the audio.
From here, the temporal entropy of the signal is computed using

\begin{equation}
H_{t}=-\sum_{t=1}^{n}A(t)log(A(t))log(n^{-1})
\end{equation}

Note that Sueur et al. \cite{Sueur_2008} use a base 2 log, but any log base gives the same result because of the change of base law.

\textbf{Spectral Entropy} \cite{Sueur_2008}: Measures the dispersion of signal over frequency. A high spectral entropy indicates that a signal is more ‘noisy’ whereas signals with lower spectral entropy sound more like pure tones. A probability mass function is computed, this time using STFT coefficients rather than the analytic signal. These are used to compute the spectral entropy using the same equation to temporal entropy, only that frequency components are considered instead of amplitude. The spectral entropy is computed for each window in the STFT, and these are averaged out to give one value. This can be split per frequency to give the spectral entropy for a specific frequency band. This could be determinstic of cicada choruses in particular, which because they a low overall spectral entropy due to a high amount of energy is in small frequency ranges where they call.

\textbf{Background Noise} \cite{Towsey_2013}: Measures the amplitude of the stationary component of the audio, where volume is approximately constant. For the constant cicada choruses, background noise will be higher than average, as the cicada signal is a constant signal, and hence will be detected by this approach to be noise. Additionally, heavy rain will also have a higher background noise than average because the number of audible raindrops per second will be high enough that there will be a stationary noise component.

This is calculated by computing the amplitude envelope of the audio signal and placing samples of the envelope into a 100-bin histogram. From this, the background noise is equal to $Mo+\delta$ where $Mo$ is the mode and $\delta$ is the standard deviation of the distribution.

The standard deviation of the background noise can itself also be used as a spectral index. Higher standard deviations indicate more variability in the background noise (or possibly more signal if enough of the signal is non-stationary). This is different to temporal entropy which gives more bias towards more intense parts of the signal. For example, a relatively stationary signal with one short volume spike will have a higher relative spectral entropy compared to the standard deviation of the background noise.

\textbf{Power Spectral Density (PSD)} \cite{Welch_1967}: Measures the average power of the signal over selected frequency regions. This is done by averaging the magnitude of STFTs for a specific frequency region. The specificity of the frequency region can target different types of sounds. Because some sounds are louder than others in some frequency bands, this could be characteristic of some sounds, such as cicadas.

\textbf{Signal to Noise Ratio (SNR)} \cite{Bedoya_2017}: Measures the ratio between the intensities of the stationary (constant amplitude) signal (noise) and the non-stationary signal (signal). There are many ways this can be calculated, but here, an approach by Bedoya et al. \cite{Bedoya_2017} based on the power spectral density is used, as this can be focussed on specific frequency ranges, and has been used successfully to detect rain. This approach uses the inverse of the coefficient of variation of power spectral densities of frequencies within a given range. This is an unusual definition for the signal to noise ratio for audio, because noise in audio is uniformly spread across frequencies, whereas signals are more concentrated on specific frequencies. As such, under Bedoya et al's. definition, Gaussian white noise has a very high signal to noise ratio. Nonetheless, this feature will identify rain, which has intensity somewhat uniformly spread across frequencies.

\textbf{Intensity-Based Signal to Noise Ratio (ISNR)}: We slightly modify Bedoya et al's. \cite{Bedoya_2017} signal to noise ratio to measure variations in intensity rather than frequency. Here, we define the intensity signal to noise ratio to be the coefficient of variation of the intensity over some time period. The intensity is calculated by summing the magnitudes of the frequency components of the STFTs for each window and taking the average over the entire file. The standard deviation is then calculated using the differences between each frame and the mean intensity. This actually measures variation in volume, but if we define the noise here to only represent the stationary component of the noise, then this is analogous to the true signal to noise ratio, because high fluctuations in volume will be caused by non-stationary sources, such as bird calls, but also rain drops.

\textbf{Segmental Signal to Noise Ratio (SSNR)} \cite{Hansen_1998}: The intensity-based signal to noise ratio is calculated to 0.1 second segments and average out. This could better indicate average signal to noise ratio in the recordings. Because the signal to noise ratio has been found to be problematic in evaluating the clarity of human speech, the segmental signal to noise ratio is used \cite{Hansen_1998}. This could carry over to bioacoustic analysis, to better estimate how noisy a recording is.

\textbf{Acoustic Complexity Index (ACI)} \cite{Pieretti_2011}: Measures the average absolute difference of the intensities of two consecutive STFT windows of the signal per frequency. Frequency bands (or the entire spectrum) can be averaged out to give ACIs for different frequency ranges. The ACI for a given frequency band is:

\begin{equation}
ACI=\frac{\sum_{k=1}^{n}|i_{k}-i_{k-1}|}{\sum_{k=1}^{n}I_{k}}
\end{equation}

Where $i_{k}$ is the intensity of a given frequency band for a time segment $k$, and $n$ is the total amount of time being examined.

\textbf{Spectral Cover (CVR)}\cite{Towsey_2014a,Ferroudj_2015}: Measures the fraction of cells in a spectrogram that exceed an intensity threshold. This can be examined in terms of frequency ranges (the mid-band activity described by Towsey et al. \cite{Towsey_2014a} is essentially this). In our approach, we directly use the modulus of the Fourier Transform coefficients and use two separate empirically derived thresholds which indicate low and medium intensities: 0.0001 and 0.0003.

\subsubsection{Mel Frequency Cepstral Coefficients (MFCCs)} 

Alongside acoustic indices, we also use MFCCs as part of our feature set. These measure the energy of frequency components in the mel spectrum. This is used to better represent how humans perceive sound. Converting from the linear frequency spectrum to the mel spectrum is done using:

\begin{equation}
m=2595log_{10}(1+\dfrac{f}{700})
\end{equation}

or, alternatively

\begin{equation}
m=1127ln(1+\dfrac{f}{700})
\end{equation}

where $m$ is the frequency in mels, and $f$ is the freqeucny in Hz \cite{Vergin_1999}.

 The MFCCs are calculated by applying a mel filterbank to the power spectrum and summing the energies of each filter. The logarithm, followed by the discrete cosine transform of each frequency component is then taken to give the coefficients. In this analysis, we take 33 coefficients are taken between 0-11.025 kHz. The number 33 was chosen because the implementation of the Fast Discrete Cosine Transform used required a value of $2^{n}+1$ coefficients. First and second deltas are also used. These are derived using
 
 \begin{equation}
 \delta M_{n}=M_{n+1}-M_{n-1}
 \end{equation}
 
 \begin{equation}
 \delta \delta M_{n}=\delta M_{n+1}-\delta M_{n-1}
 \end{equation}
 
 where $M$ is the $n$'th MFCC, $\delta M_{n}$ is the $n$'th first delta, and $\delta \delta M_{n}$ is the $n$'th second delta.

Because rain is noisy, unstructured, and contains short intensity peaks, it is not expected that MFCCs will classify rain well, because MFCCs are known to have problems with similar sounds \cite{Chu_2008,Chu_2009}. Nonetheless, MFCCs will be tested as a baseline comparison to other acoustic indices. Additionally, MFCCs are likely to classify cicada choruses accurately because of their structured and constant sounds.

To simplify subsequent sections in this work, MFCCs will not be considered to be acoustic indices.

\subsubsection{Feature Sets}

In this evaluation, we test seven feature sets. Abbreviated forms are shown in subsequent tables:

\begin{itemize}
	\item All acoustic indices \textit{(Indices)}.
	\item All acoustic indices, and acoustic indices for specific frequency ranges (0--500 Hz, 500--1 kHz, 1--3 kHz, 3--5 kHz, 5--7 kHz, 7--9 kHz, 9--11 kHz) where applicable, i.e. ACI, Spectral Entropy, SNR, ISNR, SSNR, PSD, CVR (both thresholds) \textit{(FreqIndices)}.
	\item All MFCCs \textit{(MFCCs)}.
	\item All MFCCs, excluding delta \textit{(MFCCSNo$\delta$)}.
	\item All MFCCs, acoustic indices, and acoustic indices for specific frequency ranges \textit{(All)}.
	\item All MFCCs, acoustic indices, and acoustic indices for specific frequency ranges, but excluding MFCC deltas \textit{(AllNo$\delta$)}.
	\item Subset of attributes using the Corellation Feature Subset Selection algorithm of Hall \cite{Hall_1998} with greedy backward searching, which selects for both predictive ability and lack of redundancy. This is chosen because the entire feature set contains 143 attributes for high-pass filtered sets, and 163 for other sets, many of which are near redundant or not characteristic of rain and cicadas (e.g. higher-level second delta MFCCs). This is evaluated using the WEKA \cite{Hall_2009} implementation (under the names CfsSubsetEval annd GreedyStepwise) \textit{(CFSSubset)}. This feature set was derived for each of the folds during 10-folds cross validation.
\end{itemize}

\subsection{Pre-Processing}
\label{sec:noiseReduction}

In our testing, we investigate the effects of noise filtering on our classification accuracy. While noise from rain and cicada choruses are non-stationary (i.e. not constant), we can still reduce some noise prior to classification. For our analysis, we are concerned with removing noise from recordings containing bird sound, so any sound not from a bird is considered noise. Because birds generally do not make sounds below 1 kHz \cite{Pijanowski_2011}, we can apply a 1 kHz high-pass filter to remove noise sources, such as wind and engines (i.e. from cars, planes, etc.) below 1 kHz.

Another noise reduction technique we previously found to be viable for reducing stationary background noise in bird acoustic analysis \cite{Brown_2017} is the Minimum Mean Square Error Short-Time Spectral Amplitude Estimator (MMSE STSA) \cite{Ephraim_1984}. We test this approach and high-pass filtering here to determine if they improve rain and cicada classification accuracy.

These filters change the audio signal, which consequently changes the values of the acoustic features derived by the system, which in turn potentially changes classifier performance. 

Because high-pass filtering attenuates frequencies between 0--1 kHz, acoustic features focusing on frequencies in this region. MFCCs are also calculated between 1--11.025 kHz when high-pass filtering is used (33 coefficients are still calculated).

\subsection{Evaluation Measures}

While rain classifiers and filters could be utilised in a range of bioacoustic analysis tasks, the amount of rain that users want to remove will vary depending on the application. Some users might want to only keep the cleanest of signals, whereas some might go to great lengths to extract bird sounds from recordings with very high noise interference, and want to be sure to keep everything that might contain a bird sound. Because of this, we utilise a threshold-based approach where users can change the sensitivity of the classifier to suit their needs. The threshold we use is the prediction probability specified by the classifier.

For threshold-based classification, Area Under the Curve (AUC) is a more relevant accuracy measure than the number of correct predictions. This is based on examining the area under a Receiver Operating Characteristic (ROC) curve, which plots how the false positive rate and true positive rates change as the prediction threshold changes. A higher AUC will maintain classification accuracy better over all thresholds, rather than at one threshold. Because of this, we measure and select classification configurations with the highest AUC. Accuracies (=$\frac{\text{Correct}}{\text{Total}}$) are also given as an easily understandable measure, if slightly misleading in our context.

We evaluate a large number of combinations of noise reduction approaches, feature sets, and classifiers, using 10-fold cross validation. We use the same seed for randomly generating the folds for each configuration to ensure a fair test. The combinations with the highest AUC for both rain and cicada chorus detection are selected to be our best classifiers.

\subsection{Evaluation Data}
\label{sec:evaluationData}
Recording data used in this evaluation was collected by the Samford Ecological Research Facility (SERF) in October 2010. This data contains a diverse range of sounds from many sources including many birds and insects. It contains prominent cicada choruses and heavy rain during some periods. This group has used their recordings in several publications previously \cite{Towsey_2014a,Truskinger_2014,Ferroudj_2014}.

A selection of half-hour long chunks of audio from four day-long recordings are collected, totalling 24 hours of audio for testing. These are collected randomly from different times of the day, with roughly two hour gaps between chunks. Samples containing rain and cicadas are ensured to be in the sample data. This large selection of samples is split into 10-second long chunks. A smaller, randomly selected collection of samples, amounting to approximately 100 minutes of audio, is collected from the larger set is used for training data for the model. These samples are manually labelled by us as either containing or not containing the noise interference sources being examined. The sample set is reduced to mono and downsampled to 22.05 kHz to decrease computation time. This will not remove bird sounds, as birds typically do not call above 11 kHz \cite{Pijanowski_2011}.

\section{Results and Discussion}
\label{sec:raincicadaresults}

\subsection{Rain Classification}

We perform the classification accuracy tests, sorting by area under the curve. The top 10 configurations are shown in Table \ref{tab:bestRainAUC}. There are several features to be dervied from this test:

\begin{itemize}
	\item SMO, Random Forests, and IBk all classify rain better than the other classifiers. This trend continues beyond the top 10.
	\item Using MFCCs on their own produces weaker results for classifying rain, compared to acoustic indices or combinations of acoustic indices and MFCCs (the best configuration using only MFCCs is the 10\textsuperscript{th} best overall with an AUC of 0.9817).
	\item The MMSE STSA filter is not useful for rain classification, and actually hinders accuracy. This is likely because stationary noise components of the rain (such as distant rainfall) are removed.
	\item The differences in AUC between top configurations are very small and are probably prone to random error through changing the seeds used for cross validation and to generate classification models. The best configuration cannot be definitively stated. We nonetheless select the top configuration here for the purposes for further testing and reporting results.
	\begin{itemize}
	\item This error is difficult to calculate and Weka does not provide it. Repeating cross-validation and giving the error that way does not give an accurate error bound for the AUC because cross-validation already implies repeated tests, and thresholds with the highest accuracy might change between runs.
	\end{itemize}
\end{itemize}

\begin{table}[h]
	\caption{Rain classification configuration results sorted by AUC (brackets after IBk represents value for $k$)}
	\label{tab:bestRainAUC}
	\centering
	\footnotesize
	\begin{tabular}{cccccc}
		\toprule
		HPF & MMSE & Feature Set & Classifier & AUC & Accuracy\\
		\midrule
		Yes & No & All & RF & 0.9881 & 95.8\%\\
		\midrule
		Yes & No & AllNo$\delta$ & RF & 0.9881 & 95.9\%\\
		\midrule
		Yes & No & CFSSubset & RF & 0.9878 & 95.6\% \\
		\midrule
		Yes & No & AllNo$\delta$ & SMO & 0.9877 & 95.6\%\\
		\midrule
		Yes & No & FreqIndices & RF & 0.9875 & 95.6\%\\
		\midrule
		No & No & FreqIndices & RF & 0.9873 & 95.8\%\\
		\midrule
		Yes & No & FreqIndices & SMO & 0.9871 & 95.4\%\\
		\midrule
		Yes & No & All & IBk(17) & 0.9870 & 95.3\%\\
		\midrule
		Yes & No & All & SMO & 0.9869 & 95.3\%\\
		\midrule
		Yes & No & AllNo$\delta$ & IBk(21) & 0.9868 & 95.5\%\\
		\bottomrule
	\end{tabular}
\end{table}

 This compares favourably with prior research. Ferroudj et al's. \cite{Ferroudj_2014} filter achieves an AUC of 0.8660. For Bedoya et al's. \cite{Bedoya_2017} threshold technique, we vary the PSD and SNR thresholds to determine the AUC. After an initial sensitivity analysis we set the threshold $y$ of the PSD for each step $x$ to be $y(x)=3\times10^{-5}x^{2}-3\times10^{-5}x$ and the SNR threshold $z$ for each step $x$ is set to be $z(x)=0.64+0.01x$. Doing this gives an AUC of 0.8691. This is almost certainly not optimal, but it is doubtful that it would be possible to achieve better results than the best used in our analysis.
 
 Note that the accuracy of these classifiers is taken from their default threshold. This is not always the threshold with the highest accuracy. For example, with the top configuration, an accuracy of 96.3\% is possible by simply changing the threshold to 0.6583 (or 95.9\% at a threshold of 0.1728).

The subset used by the best configuration utilised two acoustic indices covering the whole spectrum, 23 acoustic indices with specific frequency bounds, 2 MFCCs, 2 first order delta MFCCs, and 4 second order delta MFCCs. All frequency ranges are represented and all acoustic indices are represented except for background noise, temporal entropy, and the intensity-based SNR for at least some frequency ranges.

\begin{figure}[h]
	\centering
	\includegraphics[width=1\linewidth]{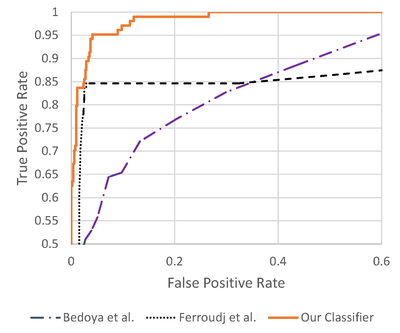}
	\caption{ROC curve comparing existing apporaches from literature. Axes are cropped to better show contrast.}
	\label{fig:roc}
\end{figure}

The value of using AUC can be clearly shown by comparing the best configuration in terms of AUC to the one of the best configurations in terms of accuracy. Figure \ref{fig:roc} shows the configuration with the highest AUC compared with curves generated by using the methods of Ferroudj et al. \cite{Ferroudj_2014} and Bedoya et al. \cite{Bedoya_2017}.

The approaches from previous research are less accurate at all thresholds. Note that the thresholds Ferroudj et al's. \cite{Ferroudj_2014} approach are based on changing the outcome of some rules within the classification tree, rather than changing the thresholds of each individual rule. Because of this, the curve is fairly flat, despite the fact that the classifier performs about as well as ours at its default threshold. However, as previously stated, Bedoya et al's. \cite{Bedoya_2017} is based on varying two thresholds. Note that there are some benefits to using Bedoya et al's \cite{Bedoya_2017} approach over ours, because its thresholds correlate with intensity, whereas ours do not, instead correlating with the likelihood that a recording contains rain that irreparably damages any bird signal in a recording.

We applied our classifier with the highest AUC to the larger 24 hour dataset. On manual inspection, heavy rain is almost always classified correctly at higher thresholds, with only some occasional barely audible bird calls appearing. Notably, these samples with bird calls are given lower prediction probabilities than samples from a similar time in the recording, with rainfall of a similar intensity but no bird calls, even if they are higher than ideal.

\begin{figure}[h]
	\centering
	\includegraphics[width=1\linewidth]{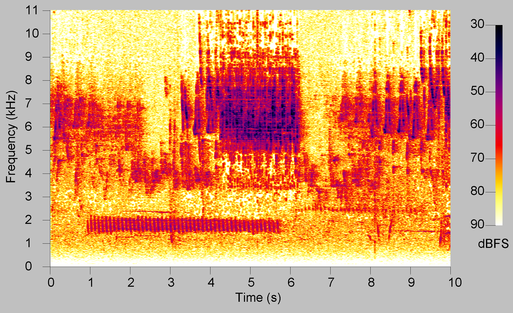}
	\caption{White-browed Scrubwren call misclassified as rain (prediction probability=0.71)}
	\label{fig:badrain}
\end{figure}

Egregious misclassifications occur infreqeuntly. One example is very prominent White-browed Scrubwren calls in samples containing no rain being classified as rain. These calls are loud, cover a somewhat broad frequency range (5--8 kHz), rapidly changing in pitch (see Figure \ref{fig:badrain}). These are sometimes classified as rain even at high thresholds, so these errors are difficult to avoid. The only other egregious misclassifications we could find at somewhat high thresholds were a result of the sounds of researchers' footsteps very close to the microphone, which are not in the training data.

\begin{figure}[h]
	\centering
	\includegraphics[width=1\linewidth]{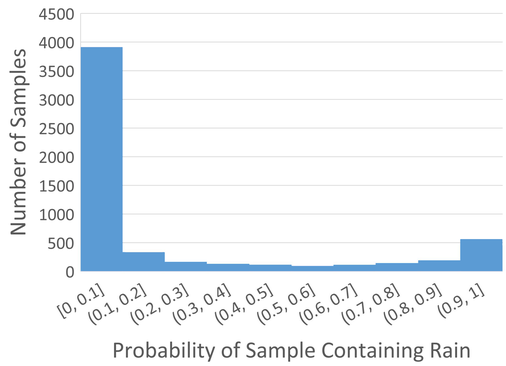}
	\caption{Distribution of predicted rain samples in the full dataset}
	\label{fig:rainhist}
\end{figure}

Figure \ref{fig:rainhist} shows how many 10-second samples in the larger set were classified as containing rain with different probabilities. Clearly, the classifier is confident in the majority of cases with its classification, with few samples having a probability of between 0.1--0.9 of containing rain, according to the classifier.

\subsection{Cicada Classification}

Table \ref{tab:bestCicada} shows the best configurations for cicada chorus classification. This shows that this classification task can be carried out almost perfectly by many classification configurations. The choice of classifier appears to be the most significant factor, with SMO, Random Forests, and particularly IBk producing the best results. Note that, accuracies are given at a threshold of 0.5, which is not always optimal, hence, a configuration can have an AUC of 1.0000 without achieving 100\% accuracy.

\begin{table}
	\caption{Cicada classification configuration results sorted by AUC (brackets after IBk represents value for $k$)}
	\label{tab:bestCicada}
	\centering
	\footnotesize
	\begin{tabular}{cccccc}
		\toprule
		HPF & MMSE & Feature Set & Classifier & AUC & Accuracy\\
		\midrule
		Yes & No & MFCCsNo$\delta$ & IBk(1) & 1 & 100.0\% \\
		\midrule
		No & No & All & IBk(1) & 1 & 100.0\% \\
		\midrule
		Yes & No & CFSSubset & IBk(3) & 1 & 99.8\% \\
		\midrule
		No & Yes & All & IBk(3) & 1 & 99.8\% \\
		\midrule
		No & No & All & RF & 1 & 99.7\% \\
		\midrule
		No & No & FreqIndices & RF & 1 & 99.5\% \\
		\midrule
		Yes & No & All & RF & $>$0.9999 & 99.8\% \\
		\midrule
		No & No & All & SMO & $>$0.9999 & 99.8\% \\
		\midrule
		No & No & AllNo$\delta$ & RF & $>$0.9999 & 99.8\% \\
		\midrule
		Yes & No & All & SMO & $>$0.9999 & 99.7\% \\
		\midrule
		No & No & MFCCs & IBk(3) & $>$0.9999 & 99.7\% \\
		\bottomrule
	\end{tabular}
\end{table}

This improves significantly on the approach by Ferroudj et al \cite{Ferroudj_2014}, whose classifier achieves 97.9\% accuracy with an AUC of 0.9531 using their feature set on our training data (the higher spectral cover threshold was used, because it was found to give better accuracy and AUC). Using only background noise and spectral entropy, as done by Towsey et al. \cite{Towsey_2014a} proved to be no more accurate than simply labelling all instances as not containing cicada choruses. This might have not been the case if half of the cicada chorus recordings in the training set also contained rain interference, reducing the value of spectral entropy as an acoustic index. Applying a random forest classifier to the two indices achieved an AUC of 0.8726, although almost all other classifiers had an AUC very close to 0.5.

Applying the cicada chorus classifier to the larger dataset shows that the cicada filter is not 100\% accurate at any threshold, but is very close, only missing a small number of samples where heavy rain almost completely overwhelms the cicada chorus in volume. This is of no consequence if the rain filter is also being applied because it will remove them anyway.

\begin{figure}[h]
	\centering
	\includegraphics[width=1\linewidth]{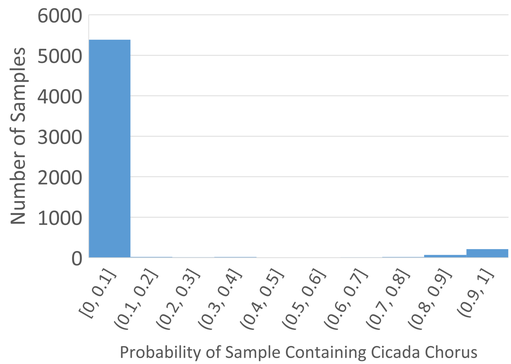}
	\caption{Distribution of predicted cicada chorus samples in the full dataset}
	\label{fig:cicadahist}
\end{figure}

Although the classification is already accurate enough such that thresholding is of limited value, we nonetheless apply a thresholding approach to examine the effects of thresholds on cicada detection. We use a Random Forest classifier (no high-pass filtering, all features) rather than IBk, because, when k=1, the classifier does not give a continuous probability distribution. Figure \ref{fig:cicadahist} shows that almost all samples (98\%) either have a probability of less than 0.1, or greater than 0.8. The first false positive we found had a prediction probability of 0.2, meaning a threshold around 0.3 or higher should not result in many false positives, and samples with cicada chorus with lower probabilities are almost completely overpowered by heavy rain.

\section{Cicada Chorus Filter}
\label{sec:cicRemoval}

\begin{figure}[h]
	\centering
	\includegraphics[width=1\linewidth]{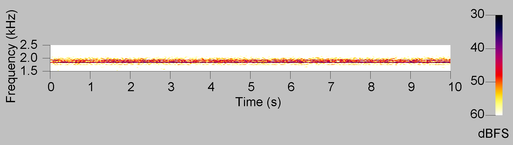}
	\caption{Cropped spectrogram showing the frequency region containing cicada sounds, demonstrating they are non-stationary. Dynamic range is reduced compared to other spectrograms in this work to emphasise contrast}
	\label{fig:croppedcicadaspectrogram}
\end{figure}

Cicada choruses are typically very loud and occupy a narrow frequency band. While a regular noise filter, such as the MMSE STSA filter, will reduce the cicada noise somewhat, it is not aggressive enough to remove the cicada sound completely because, while a cicada chorus sounds stationary when listening, it is actually non-stationary in reality. This is shown in Figure \ref{fig:croppedcicadaspectrogram} which only looks at the frequency region containing cicadas, showing that intensity does indeed fluctuate. 

A more aggressive approach to removing the cicada sounds in the recording is to completely remove the frequencies containing the cicada sounds, via the use of bandpass filtering. Most band-pass filters feature a roll-off, where frequencies are reduced less the closer they are to the cut-off frequency. However, for the cicada filter, the sinc filter is used, which has a far more aggressive roll-off compared to typical high pass filters, such that almost no information from the frequencies with cicada choruses is preserved.

The probability mass function (PMF), which calculates what proportion of energy is in a discrete frequency variable (see Section \ref{Indices}), can be used to find to a reasonable level of accuracy what frequency band the cicada is occupying. A novel algorithm is proposed using the PMF as a basis (Note: $RSD$ = relative standard deviation):

\begin{algorithm}
	\begin{algorithmic}
		\caption{Cicada filter algorithm}
		\label{alg:cicFilter}
		\FORALL{samples}
		\STATE Compute FFT
		\STATE Perform cicada detection
		\ENDFOR
		\IF{sample is detected as containing cicada}
		\FORALL{FFT Windows}
		\STATE Compute PMF
		\ENDFOR
		\FORALL{FFT Frequency Components $x$}
		\STATE Calculate the $\overline{PMF_{x}}$ and $RSD_{\overline{PMF_{x}}}$
		\ENDFOR
		\STATE Find range of frequency components with the highest $\sum\overline{PMF}$ by which consecutive frequency components $x$ obey the rule:
		 \STATE $\overline{PMF_{x}}$ $>$ 0.0125 and $RSD_{\overline{PMF_{x}}}$ $<$ 70\%.
		\STATE Apply a sinc-filter to remove this frequency range
		\ENDIF
	\end{algorithmic}
\end{algorithm}

\begin{figure*}
	\centering
	\subfloat[Raw]{ \includegraphics[width=0.5\textwidth]{"CicadaChorusSpectrogram"} \label{cicRaw} }
	\subfloat[MMSE STSA]{ \includegraphics[width=0.5\textwidth]{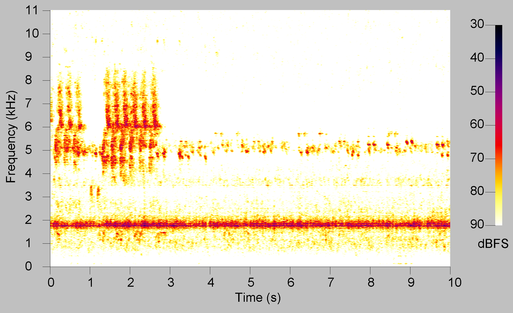} \label{cicMMSE} }\\
	\subfloat[Cicada Filter]{ \includegraphics[width=0.5\textwidth]{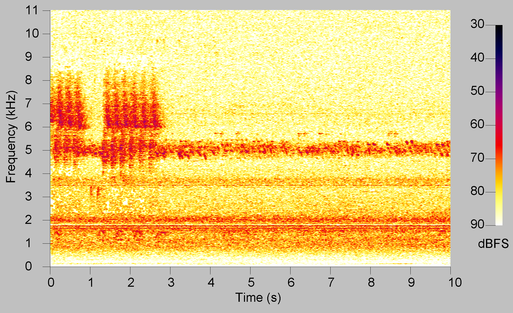} \label{cicSinc} }
	\subfloat[Cicada Filter and MMSE STSA]{ \includegraphics[width=0.5\textwidth]{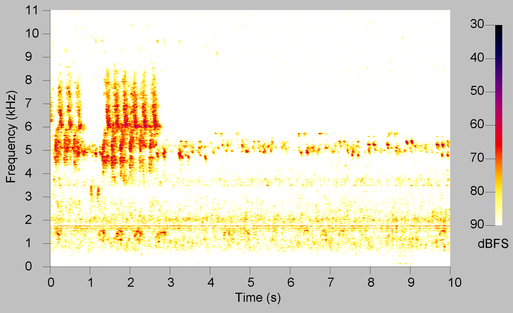} \label{cicBoth} }\\
	\subfloat[Waveform]{\includegraphics[width=1\linewidth]{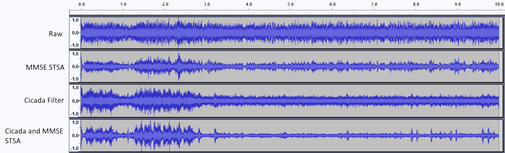}
	\label{fig:cicadafilterwaveform}}
	\caption{Effect of the cicada filter on the audio, in the form of spectrograms in the waveform. All  waveforms have been amplified individually to 0 dBFS for clarity}
	\label{fig:cicFilter} 
\end{figure*}

The effect of the cicada filter is shown in Figure \ref{fig:cicFilter}. This successfully eliminates all of the cicada sound, albeit at the expense of other sounds in the same frequency region. There is an assumption here that cicada sounds overpower all other sounds and so these are unrecoverable, which generally holds true in the dataset used. This filter makes it easier for humans to listen to recordings, as well as for automated methods to accurately process the data.

\subsection{Evaluation and Discussion of Cicada Chorus Filter}
\label{sec:cicFilterResults}

To evaluate cicada filter's effectiveness, we apply the cicada filter to a 30-minute section of audio, split into 10-second chunks, known to contain cicada chorus and no rain. For this test, the cicada filter was applied to each file without testing the detection algorithm, which was tested previously. Four versions of each file are created: a raw, unprocessed audio file, an MMSE STSA filtered audio file, a cicada filtered file, and a file that has been both cicada filtered and MMSE STSA filtered. 

 To give an objective measure of the cicada filter's effectiveness, we estimate the signal to noise ratio for all versions of the files. A 1 kHz high-pass filter is also applied to each file to negate the effects of other potential non-stationary interference sources such as wind and engine sounds. The cicada filter is applied first, than the MMSE STSA, and finally the high-pass filter, with filters being skipped when not included in a set. 
 
\begin{figure}[h]
	\centering
	\includegraphics[width=1\linewidth]{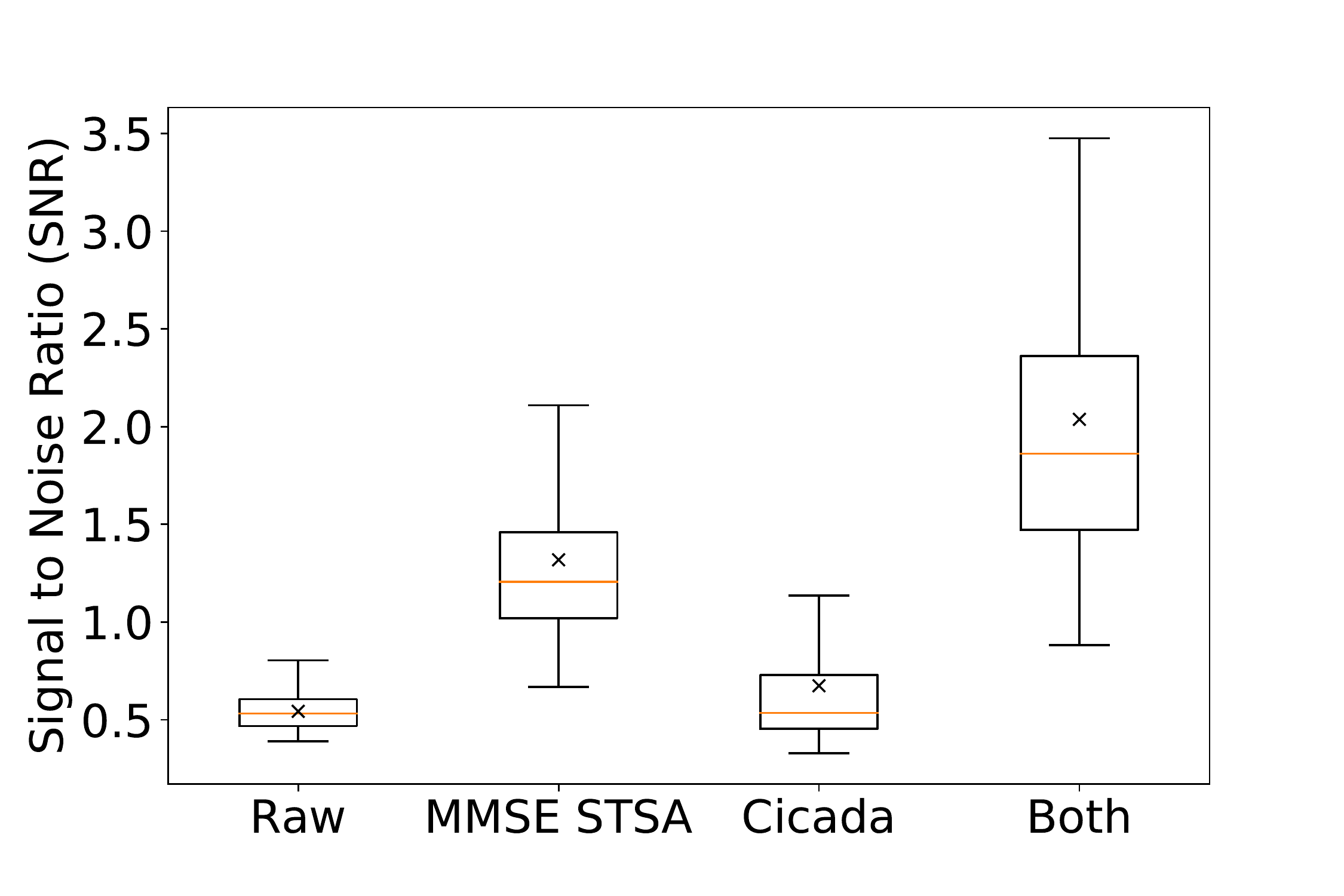}
	\caption[Effect of the Cicada filter on the estimated ISNR of the audio]{Effect of the Cicada filter on the estimated ISNR of the audio. Outliers have been removed to reduce the range of the plot to show differences. `x' indicates the mean. Samples were all high-pass filtered (1 kHz)}
	\label{fig:SNRCicadaFilter}
\end{figure}

 The intensity-based signal to noise ratio is used here for comparison because it best represents the difference between the approximately stationary (i.e. constant in volume) cicada chorus and the non-constant bird sounds. The results, shown in Figure \ref{fig:SNRCicadaFilter}, and Table \ref{tab:SNRStats} clearly show that the combination of the cicada filter and the MMSE STSA filter greatly increase the signal to noise ratio of the audio, and are more effective together than on their own.

\begin{table}
	\centering
	\caption{Effect of filters on the estimated ISNR}
	\label{tab:SNRStats}
	\begin{tabular}{cccc}
		\toprule
		Filter & Median & Quartile 1 & Quartile 3 \\
		\midrule
		Raw & 0.532 & 0.469 & 0.606 \\
		\midrule
		MMSE STSA & 1.206 & 1.019 & 1.460 \\
		\midrule
		Cicada Filter & 0.535 & 0.454 & 0.729 \\
		\midrule
		Both & 1.862 & 1.472 & 2.361 \\
		\bottomrule
	\end{tabular}
\end{table} 

This was verified by performing a two-tailed Mann-Whitney U test on all sets. The results of this test are shown in Table \ref{tab:MWU}. Every set has statistically significant differences in ISNR values, to a 99\% confidence interval except for the raw audio and cicada filter. The simliarity between the ISNRs of raw, and cicada filtered audio is suspected to be caused by the fact that cicada choruses themselves fluctuates in volume, increasing the ISNR. This is supported by the fact that the ISNRs of cicada filtered samples have a notably higher IQR compared to raw samples: When there are no other loud bird calls, the cicada chorus has a more significant contribution to the ISNR, but when other bird calls are present, they contribute more to the ISNR, and this added ISNR is more pronounced when cicada filtering has occurred.

\begin{table}
	\footnotesize
	\centering
	\caption{Mann-Whitney U Test Results for each filter against each other (values are $U$,$p$. MMSE = MMSE STSA. 180 values tested, $U$=16290 where distributions are equal.)}
	\label{tab:MWU}
	\begin{tabular}{ccccc}
		\toprule
		Filter & Raw & MMSE & Cicada & Both \\
		\midrule
		Raw & - & 67, $<$0.001 & 14793, 0.090 & 0, $<$0.001 \\
		\midrule
		MMSE& - & -& 2208, $<$0.001 & 5155, $<$0.001 \\
		\midrule
		Cicada & - & - & - & 582,$<$0.001 \\
		\bottomrule
	\end{tabular}
\end{table}

\section{Conclusions and Future Directions}
\label{sec:conclusion}

In this work, we investigated the use of acoustic indices with MFCCs in combination of classification algorithms for reducing noise from non-stationary sources (rain and cicada choruses) in bioacoustic recordings. We evaluated many configurations with different acoustic features, classifiers, and filters to detect cicada choruses and rain in bioacoustic recordings. Using an SMO classifier and applying a high-pass filter, we were able to build a classifier with an AUC of 0.9880. This classifier can have thresholds varied to change its sensitivity depending on the application. Although there were some samples obviously not containing rain with high classification probabilities, we found that overall, the classifier appeared to perform equally as well on the larger test data compared to the training data.

 We also built a cicada chorus classifier that detected cicada chorus samples with 100\% accuracy with an AUC of 1.0000 using 10-fold cross validation with the training set. Accuracy continues to very high even when the model is applied to a larger  testing set.

We also designed a filter to remove cicada choruses from affected audio by removing the frequency band where cicada choruses are most prominent. This filter accurately detected and removed a frequency range containing a cicada chorus without removing other animal calls 100\% of the time in our testing. We found that this improved the signal to noise ratio of the audio signal, suggesting that sounds outside of the cicada chorus were heard more clearly.

In future work, we wish to remove noise from more sources in bioacoustic recordings. We also want to build filters that can remove individual rain drops from recordings with lighter rain, and perhaps pursue approaches to recover information damaged by rain, rather than completely delete recordings with rain. 

Another possible addition to this work would be to build another classifier focussed on determining the intensity of rain in a sample, rather than the extent by which it interferes with clean signals, much like Bedoya et al. \cite{Bedoya_2017}, but potentially more accurate.

\section*{Acknowledgements}

We thank the Samford Ecological Research Facility (SERF) for providing us with environmental recordings used as sample data for this work. This research did not receive any specific grant from funding agencies in the public, commercial, or
not-for-profit sectors.

\section*{References}

\bibliography{bibfile}
\end{document}